\def\<~{\raisebox{-3.5pt}{\,$\stackrel{\raisebox{-3pt}{$\textstyle<$}}{\sim}$}\,}

\documentclass[amsmath,amssymb,aps,floatfix]{revtex4}

\usepackage{epsfig}
\usepackage{graphics}	% Include figure files
\usepackage{bm}		% bold math

\begin{document}

\title{Comment on `Detailed balance has a counterpart in non-equilibrium steady states'}

%\author{Aditi Simha and R. M. L. Evans}
%\affiliation{School of Physics and Astronomy, University of Leeds, LS2 9JT, U.K.}

\author{Aditi Simha\footnote{Present address: Indian Institute of Technology, Madras, Chennai, India} \& R. M. L. Evans} 
\affiliation{School of Physics and Astronomy, University of Leeds, LS2 9JT, U.K.}

\date{24th June 2009}

\begin{abstract}
Transition rates in continuously driven steady states (relevant to sheared complex fluids) were derived in \cite{JPhysA,PRL} by demanding that no information other than the microscopic laws of motion and the macroscopic observables of the system be used to describe it. This implies that the ({\em nonequilibrium}) reservoir, to which the system is weakly coupled, is fully characterised by its mean energy and mean flux. While we expect the resulting prescription for the rates in continuous- and discretised-time models to be equivalent, it is not trivial to see this from the expression for the rates derived in \cite{JPhysA}. We demonstrate this equivalence for a model of activated processes solved previously for continuous time \cite{JPhysA}, thus demonstrating consistency of the theory .     
\end{abstract}

\maketitle

% \noindent{\bf Keywords}\,\,\,Non-equilibrium processes: Exact results, Stationary states; Stochastic processes (Theory) \\ \\

\begin{figure}
%\resizebox{5.5cm}{!}{\includegraphics{combfig.eps}}
  \epsfig{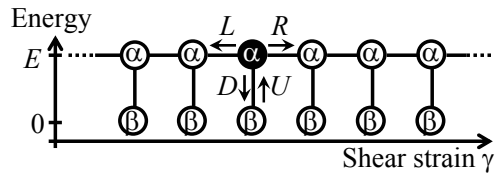}
\caption{\label{combfig} The discrete comb-shaped state-space for a simple model system studied in \cite{JPhysA} for continuous time, and here for discrete time.}
\end{figure}

The rates of stochastic transitions between microstates of a complex fluid under shear, subject to non-equilibrium noise from a reservoir of the sheared fluid, were found in \cite{JPhysA,PRL} to respect certain detailed-balance-like rules. The rules were applied to a toy model, described in \cite{JPhysA}, in which the fluid's discrete state-space has a comb-shaped connectivity of permitted transitions, as defined in Fig.~\ref{combfig}. 
This models a system whose energetically favoured states are incompatible with a large flux.
Whichever state is currently occupied by the system (that might be composed of many interacting particles), can be represented by a filled-circle, that has the appearance of a single particle hopping in a real-space landscape. The stochastic transitions occur with a certain probability in one time step.
 
As in \cite{JPhysA} (but now for discrete time), we define an equilibrium version of the model, and {\em derive} a driven-ensemble version. In the latter case, the non-equilibrium constraint is to demand that, after a long time $\tau$, the ensemble-averaged strain has a large positive value $\gamma_0$, defining a shear current $\dot{\gamma}$ by $\gamma_0\equiv \dot{\gamma}\tau$. 
As shown in Fig.~\ref{combfig}, the system can occupy states of two different types: high-energy states $\alpha$ that are connected to other $\alpha$ states, and low-energy traps $\beta$. It may seem that one is free to choose independently four parameters that define the model: the probabilities $u$, $d$, $l$, $r$ of transitions upward, downward, leftward and rightward in Fig.~\ref{combfig} respectively. At equilibrium, detailed balance requires
$l^{\rm eq} = r^{\rm eq} \equiv \omega$ 
and 
$u^{\rm eq} = d^{\rm eq}\,\exp(-E) \equiv p$
with the thermal energy scale defined by $k_B T\equiv 1$,
so that only two free parameters remain. Transition probabilities are therefore similarly restricted in the driven steady-state ensemble \cite{JPhysA}, where a transition from microstate $a$ to $b$ has probability per discrete time-step    
\begin{equation}
\label{exp}
  p_{ab}^{\rm dr}=p_{ab}^{\rm eq}\, \exp \left[\nu \gamma_{ab}
  +q_{b}(\nu) -q_{a}(\nu) -Q(\nu)\Delta t \right].
\end{equation}
Here $p_{ab}^{\rm eq}$ is the probability at equilibrium for the same transition, $\gamma_{ab}$ the net flux gained by it, $\nu$ is a parameter conjugate to the flux, and $q_{a}(\nu)$ defined for discrete transitions by 
\begin{equation}
  q_{a}(\nu)=\lim_{\tau\rightarrow \infty}
  \left[ \ln\sum_{\gamma=-\infty}^{\infty} G_{a}(\gamma,\tau)\, e^{\nu \gamma} - \tau Q(\nu) \right],
\end{equation}
a quantity measuring the system's propensity to exhibit flux in future from the initial state $a$. $Q(\nu)$ is a property of the system's steady state and is related to the average current by $dQ/d\nu = \dot{\gamma}$. 
Here, the equilibrium Green's function $G_a(\gamma,\tau)$, is the probability that the system will undergo a total strain $\gamma$ in time $\tau$, beginning from state $a$. For discrete time steps ($\Delta t=1$) Eq.~(\ref{exp}) yields for the four transitions probabilities:
\begin{eqnarray}
 r&=& \omega\, e^{\nu -Q(\nu)}\, ,\nonumber \\
 l&=& \omega\, e^{-\nu-Q(\nu)}\, , \nonumber \\
 d&=&p\,e^{E}\,e^{q_{\beta}(\nu)-q_{\alpha}(\nu)-Q(\nu)} \nonumber \\
 u&=&p\, e^{q_{\alpha}(\nu)-q_{\beta}(\nu)-Q(\nu)}\, .
\label{rate4}
\end{eqnarray}
To evaluate these we must find $q_{\alpha}(\nu)-q_{\beta}(\nu)$ and 
$Q(\nu)$. 

\begin{figure}
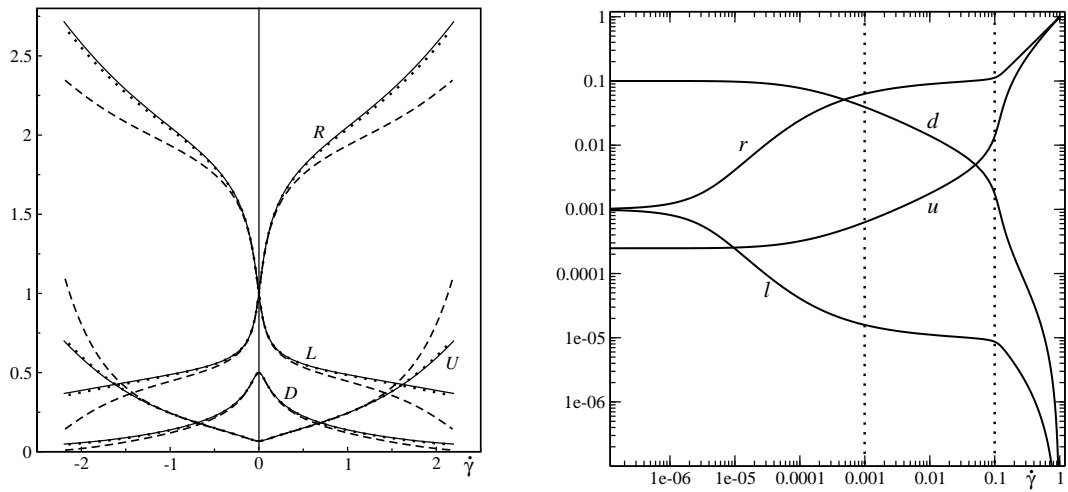

\resizebox{6.5cm}{!}{{\bf (a)} \raisebox{5mm}{\includegraphics{wcontf3.eps}}}
\qquad
\resizebox{7cm}{!}{{\bf (b)} \includegraphics{logr.eps}}
  \caption{\label{graph}(a) The transition \emph{rates} - probabilities for the discrete-time model scaled by $1/\Delta t$ - plotted for parameter values $E=2$, 
$p\,e^{E}/\omega=0.5$ for two values of  $\omega = 0.4 \,(=\Delta t)$ (dashed lines) and $ \omega=0.04= \Delta t$ (dotted lines) along with the rates of the continuous time version (solid lines).
(b) Log-log graphs of $u$, $d$, $l$ and $r$ for discrete-time hopping in the comb-shaped state-space of Fig.~\protect\ref{combfig}, as a function of mean shear flux $\dot{\gamma}$, with $E=6$, $\omega=0.001$, $p\,e^{E}=0.1$. The dotted lines separate three distinct regimes of behaviour.} 
\end{figure}

The connectivity of the state-space (Fig.~\ref{combfig}) implies relationships between the various Green's functions (simplified in \cite{PRLAB} but given here in full). A system initially in state $\beta\,$ hops to state $\alpha$ at time $t$ with probability $\,p(t)=p\,(1-p)^{t-1}$ and is governed by the propagator $G_{\alpha}(\gamma,\tau)$ from there on. This translates to
\begin{equation}
\label{propagators}
  G_{\beta}(\gamma,\tau) = \sum_{t=1}^{\tau} \,p(t)\, G_{\alpha}(\gamma, \tau-t)
  = \sum_{t=1}^{\tau}\, p\, (1-p)^{t-1}\, G_{\alpha}(\gamma, \tau-t).
\end{equation}
In terms of the quantity 
$m_{i}(\nu,\tau) \equiv \ln \sum_{\gamma=-\infty}^{\infty} G_{i}(\gamma,\tau)\, e^{\nu \gamma}$,
defined for each of the states ($i=\alpha, \beta$), Eq.~(\ref{propagators}) yields
\begin{equation}
  p\,e^{m_{\alpha}(\nu,\tau)-m_{\beta}(\nu,\tau)}
  = \left[e^{ m_{\beta}(\nu,\tau+1)- m_{\beta}(\nu,\tau)}-1+p\right].
\end{equation}
In the limit $\tau \rightarrow \infty\,$, 
$ m_{\beta}(\nu,\tau+1)- m_{\beta}(\nu,\tau)\to Q(\nu)\,$ \cite{JPhysA} and $\,m_{i}(\nu,\tau)-m_{j}(\nu,\tau) \to q_{i}(\nu)-q_{j}(\nu)\,$. Hence,
\begin{equation}
  e^{q_{\alpha}(\nu)-q_{\beta}(\nu)}=\frac{1}{p}
  \left[ e^{Q(\nu)}-1+p\right] \mbox{.}
\end{equation}
Since only the $\alpha$ states carry flux, the mean shear flux 
$\dot{\gamma} = (r-l)\,f_{\alpha} = (r-l)\,u/(u+d)$ 
is determined solely by $r$ and $l$ and the occupancy $f_{\alpha}$ of the $\alpha$ state. Using $\,dQ/d\nu=\dot{\gamma}$, this gives a differential equation for $Q\,$:
\begin{equation}
  \frac{dQ}{d\nu} = 
  \frac{2\, \omega\,\sinh\nu\, e^{Q}}{1+ p^{2}\,e^{E}\,(e^{Q}-1+p)^{-2}}
\end{equation} 
which can be solved for $Q(\nu)$. The condition $Q(0)=0$ (coming from normalization of the propagator) sets the constant of integration, yielding 
\begin{equation}
  e^{Q(\nu)} 
   = \omega\, \cosh(\nu)+1-\omega-\frac{p}{2}(e^{E}+1) 
   + \sqrt{\left(\omega \cosh(\nu) + \frac{p}{2}
  (1-e^{E})-\omega\right)^{2}+p^2\,e^{E}}\,\,\,.
\end{equation}
Eliminating $\nu$ in favour of $\dot{\gamma}$, Eq.~(\ref{rate4}), thus gives the following four simultaneously equations that uniquely specify the rates in the driven steady state: 
\begin{eqnarray}
\label{const1}
  {r - l} &=& \left(1+d/u \right) \dot{\gamma}	\\
  \omega^2\, (d\,u) &=& p^{2} e^{E}\,\, (r\,l)	\\
  u &=& 1-\sqrt{r\,l}\,(1-p)/\omega	\\
  \omega \,(r + l + d -u) &=& \sqrt{r\,l} \,\left[2\,\omega+p\,(e^{E}-1)\right]\, .
\label{const4} 
\end{eqnarray}
The quantities $r$, $l$, $d$, $u$ are dimensionless, normalized probabilities, in contrast to the dimensionful {\it rates} analysed in the continuous-time version of the model \cite{JPhysA}. 
In the limit of short time-steps (where the unit of time is redefined to be $\Delta t\to 0$), all hopping probabilities per time-step vanish in proportion to $\Delta t$. We define the transition \emph{rates} to be the corresponding probabilities scaled by $1/\Delta t\,$ (denoted by $R$, $L$, $D$, $U$\, as in \cite{JPhysA}) with units chosen such that $R^{\rm eq}=L^{\rm eq}=1$ in the continuous-time version of the model. The rates are plotted in Fig.~\ref{graph}a for parameter values consistent with Fig. (3) of \cite{JPhysA}. As $\Delta t\to 0$, our rates converge to those calculated previously for continuous time. In addition, it is easy to show that the constraints Eqs.(\ref{const1}-\ref{const4}) lead to those in \cite{JPhysA} exactly, in the limit  
$\Delta t\to 0\,$, thus confirming the self-consistency of the formalism in \cite{JPhysA}.

The constraint of normalization in the discrete-time case leads to significant quantitative differences in the model's behaviour for non-vanishing time steps in the limit of large flux.
The probabilities are plotted again in Fig.\ref{graph}b on logarithmic axes (for positive velocities), where the parameters $E=6$, $\omega=0.001$, $p\,e^{E}=0.1$ (as in \cite{JPhysA}) have been chosen to separate the relevant time-scales for clarity. Vertical lines on the figure separate three different regimes of imposed current $\dot{\gamma}\,$ (as also seen in the continuous-time version of the model \cite{JPhysA}): (a) the near-equilibrium regime at the lowest velocities where the rates of activation and relaxation continue to be governed by the equilibrium principle of detailed balance, (b) a regime where $\omega\: \<~\: \dot{\gamma}\: \<~\:p\,e^{E} $, in which $r$ and $l$ remain approximately constant with increasing $\dot{\gamma}$, while the required flux is achieved by variations of only the activation and relaxation probabilities $u$ and $d$, and (c) a regime where all four probabilities vary --- unlike the continuous-time result, the flux-inhibiting transition probabilities $l$ and $d$ strictly vanish as $\dot{\gamma}\to0$ while the favoured probabilities $r$ and $u$ rise towards unity.

In summary, we have applied the rules that were derived for transition rates in \cite{JPhysA} for a particular class of nonequilibrium steady states to a simple model system with a comb-shaped state-space which it explores in discrete time-steps. Even for such a simple system, the application of the transition-rate rules is a non-trivial procedure.
(See \cite{PRLAB} for some recent simplifications.) Having carried out that procedure, we have been able to show that in the limit of small time steps the results tend to those derived for the continuous time version of the same model thereby demonstrating the self-consistency of the formalism.   

\vspace*{3mm}
\noindent{\bf Acknowledgments}\, This work was funded by the Royal Society and by EPSRC grant GR/T24593/01.

\end{document}